\documentclass[twocolumn,showpacs,preprintnumbers,amsmath,amssymb]{revtex4}
\usepackage{graphicx}
\usepackage{textcomp}

\begin{document}
\title{Classical and quantum interference in multiband optical Bloch oscillations}
  \normalsize
\author{Stefano Longhi \footnote{Author's email address: longhi@fisi.polimi.it}
}
\address{Dipartimento di Fisica and Istituto di Fotonica e
Nanotecnologie del CNR, Politecnico di Milano, Piazza L. da Vinci
32, I-20133 Milano, Italy}

%
\bigskip
\begin{abstract}
\noindent Classical and quantum interference of light propagating in
arrays of coupled waveguides and undergoing multiband optical Bloch
oscillations (BOs) with negligible Zener tunneling  is theoretically
investigated. In particular, it is shown that Mach-Zehnder-like
interference effects spontaneously arise in multiband BOs owing to
beam splitting and subsequent beam recombination occurring in one BO
cycle. As a noteworthy example of quantum interference, we discuss
the doubling of interference fringes in photon counting rates for a
correlated photon pair undergoing two-band BOs, a phenomenon
analogous to the manifestation of the de Broglie wavelength of an
entangled biphoton state observed in quantum Mach-Zehnder
interferometry.
\end{abstract}

%

 \maketitle

\section{Introduction}
Since the pioneering works by Bloch, Zener and Wannier on the
coherent electron dynamics in biased crystalline solids
\cite{Bloch34}, it is well established that quantum particles in
periodic potentials subjected to an external force do not delocalize
but undergo a high-frequency oscillatory motion known as a Bloch
oscillation (BO). Tunneling to higher-order bands, referred to as
Zener tunneling (ZT) and observed for a strong external bias, is
responsible for BO damping and broadening of Wannier-Stark
resonances. Owing to detrimental dephasing and many-body effects,
the experimental verification of BOs failed for many decades.
Nowadays BOs are considered a rather universal wavy phenomenon and
their observation has been reported in different physical systems
for both quantum particles and classical waves. In the quantum
realm, BOs have been observed for electrons in biased semiconductor
superlattices \cite{Mendez88} and for Bose-Einstein condensates in
accelerating optical lattices \cite{BEC}. Classical analogues of
BOs, based on interference effects of classical waves in periodic
media, have been proposed and experimentally observed using either
optical \cite{Morandotti99,Sapienza03,Christodoulides03,Trompeter06}
or acoustical \cite{acoustic} waves. Optical BOs in artificial
materials or at the nanoscale have been also recently predicted
\cite{nano}. Among various photonic structures, arrays of coupled
waveguides with transverse refractive index gradients have provided
an extremely rich laboratory tool to visualize the classical wave
optic analogues of BOs
\cite{Morandotti99,Christodoulides03,Trompeter06} and related
phenomena, such as the coherent superposition of BOs and ZT (the
so-called Bloch-Zener oscillations \cite{Breid06}) occurring in
binary structures and recently observed in circularly-curved
femtosecond-laser-written waveguide arrays \cite{Dreisow09}. In
another physical context, application of Bloch-Zener oscillations to
matter-wave interferometry has been also proposed \cite{Breid07}.
While there has been a lot of interest in the propagation of light
waves in complex photonic structures like lattices and
superlattices, quasi-crystals, and even random lattices, most of
previous works focused on propagation of classical light and used
the photonic structures as 'classical simulators' of electronic or
matter wave systems in periodic or random potentials. However, it is
known that coupled waveguides behave similarly to beam splitters
(see, for instance, \cite{Lai91}) and may therefore show quantum
interference effects when probed with nonclassical light.
Demonstration of key quantum effects of nonclassical light in
silicon-based based waveguide circuits has been recently reported in
Ref.\cite{Politi08}. As shown in recent works
\cite{Longhi08,Rai09un}, nonclassical light consisting of only
particle-like quanta propagating in waveguide arrays with a
superimposed transverse refractive index gradient can also produce
optical BOs. Remarkably, in addition to classical wave Bragg
scattering the quantum nature of light introduces new quantum
interference effects. In particular, two-photon Hong-Ou-Mandel
quantum interference \cite{Hong87} has been predicted for pairs of
correlated photons undergoing Bloch-Zener oscillations in binary
arrays. Here ZT periodically mixes photons belonging to the two
minibands of the array and thus acts like a beam splitter
\cite{Longhi08}. Recently, quantum correlations of photon pairs
undergoing discrete diffraction in homogeneous arrays have been
investigated as well, and their classical counterpart has been
experimentally observed in Hanbury Brown-Twiss intensity correlation
measurements \cite{Bromberg09}. The investigation of quantum
interference effects in complex photonic structures such as
homogeneous, inhomogeneous or even random photonic lattices has two
main motivations. On the one hand, they may offer the possibility of
engineering photon entanglement and of transporting nonclassical
light \cite{Rai08,Bromberg09}; on the other hand, they enable to
utilize optical-quantum analogies in the opposite direction, i.e.
trying to observe quantum version of classical wave phenomena (such
as the analogue of the Talbot effect
in the second quantized setup \cite{Rai08}).\\
It is the aim of this work to provide a comprehensive analysis of
classical and quantum interference of light undergoing BOs in
inhomogeneous waveguide arrays in the regime of multiband excitation
and negligible ZT, i.e. in a complementary regime of that previously
considered in Refs.\cite{Breid07,Longhi08}. In spite of the absence
of interband transitions, it is shown that Mach-Zehnder-like
interference effects do spontaneously occur for a multiband
excitation of the array at the input plane. In this regime, the
input beam breaks up into two (or more) wave packets belonging to
different bands of the array which follow distinct paths in the real
space and recombine after a full BO cycle. Wave and photon
interference effects after a full BO cycle may be thus observed by
classical and nonclassical light illumination, even in absence of
ZT. In particular, it is shown that array excitation with correlated
photon pairs tilted at the Bragg angles enables to observe quantum
interference patterns in two-photon correlation measurements
analogous to those occurring in quantum Mach-Zehnder interferometry
as a manifestation of the de Broglie wavelength of
entangled photon states \cite{Rarity90,Boto00,Edamatsu02,Walther04}.\\

\section{Wave-optics model of multiband Bloch oscillations in a waveguide array}
The starting point of our analysis is provided by a rather
standard wave optics model describing BOs of monochromatic light
waves at carrier frequency $\omega=2 \pi c_0/ \lambda$ propagating
in a weakling guiding one-dimensional waveguide array [Fig.1(a)],
with a periodic refractive index profile $n(x)$ and with a
superimposed transverse refractive index gradient $Fx$
\cite{Trompeter06,Longhi08,Longhi06}. In the paraxial
approximation, the slow evolution of a scalar field component
$\phi(x,z)$  along the paraxial $z$ direction is governed by the
Schr\"{o}dinger-like wave equation (see, for instance,
\cite{Longhi06})
\begin{equation}
i \lambdabar \phi_z=- \frac{\lambdabar^2}{2 n_s}
\phi_{xx}+[V(x)-Fx] \phi,
\end{equation}
\begin{figure}[htbp]
\includegraphics[width=8.6cm]{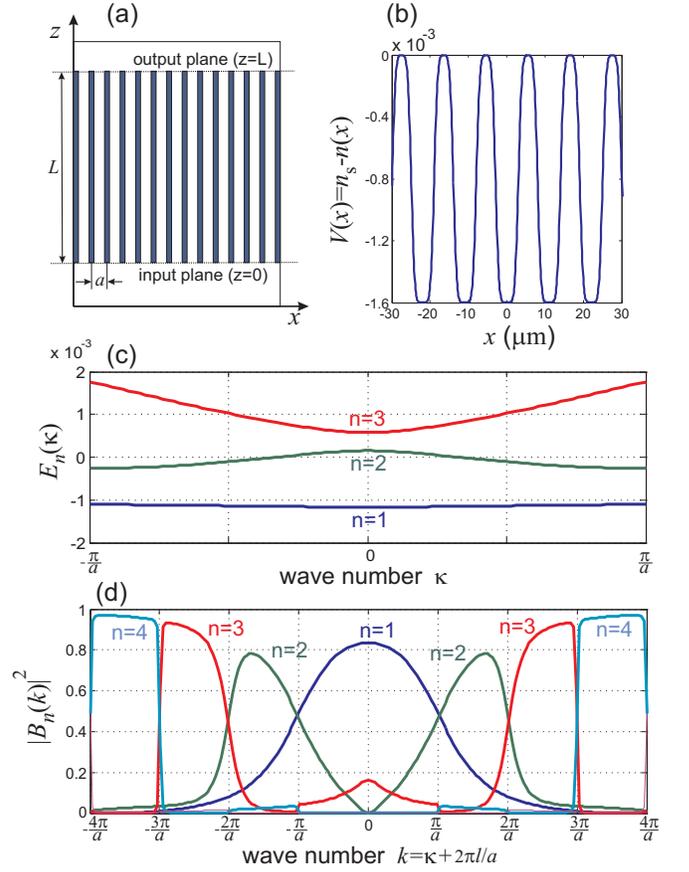}\\
   \caption{(color online) (a) Schematic of a singly-periodic one-dimensional waveguide array.
   (b) Behavior of the array refractive index $V(x)=n_s-n(x)$ used in numerical simulations. The transverse
   index ramp term is applied by e.g. transverse thermal heating or
   by circularly-curving the waveguide axis (not shown  in the figure). (c) Band diagram. (d) Behavior of the Fourier
   coefficients $|B_n(k=\kappa+2 \pi l /a)|^2=|\theta_{n,l}(\kappa)|^2$ of Bloch modes $\varphi_n(x,\kappa)$ for a few low-order bands.}\label{fig1}
\end{figure}
where $\lambdabar=\lambda/(2 \pi)$ is the reduced wavelength,
$n_s$ is the substrate refractive index, $V(x)= n_s-n(x)$ accounts
for the periodic modulation of the refractive index with spatial
periodicity $a$ [$V(x+a)=V(x)$], and $F$ is the superimposed
transverse index gradient. The array is assumed to be confined in
the region $0<z<L$ [see Fig.1(a)] and array excitation is
accomplished at the input plane $z=0$. Since $V(x)=F=0$ for $z<0$
and $z>L$, usual paraxial diffraction occurs before and after the
light beam enters and leaves the arrayed structure. The sample
length $L$ is typically chosen to be equal to the characteristic
period $z_B$ of BOs at a reference value $F=F_0$ of index
gradient, i.e. $L=z_B$ where \cite{Longhi06}
\begin{equation}
z_B=\frac{\lambda}{ F a}.
\end{equation}
In absence of the transverse index gradient ($F=0$), the modes of
the arrayed structure are the Bloch states $\varphi_n(x, \kappa)$
with corresponding dispersion curves $E_n(\kappa)$, where
$n=1,2,3,...$ is the band index, $\kappa$ is the Bloch wave number
(quasi momentum) which varies in the first Brillouin zone $-\pi/a <
\kappa \leq \pi/a$, and the orthogonal and normalization conditions
$\langle \varphi_{n'}(x,\kappa')| \varphi_{n}(x,\kappa)
\rangle=\delta_{n,n'} \delta(\kappa-\kappa')$ hold. $\varphi_n(x,
\kappa)$ and $E_n(\kappa)$ are found as eigenfunctions and
eigenvalues of the problem $\mathcal{H}_0 \varphi_n(x,
\kappa)=E_n(\kappa) \varphi_n(x, \kappa)$ with periodic Hamiltonian
$\mathcal{H}_0=-[\lambdabar^2/(2n_s)]\partial^{2}_{xx}+V(x)$.
According to the Bloch theorem, one can write $\varphi_n(x,
\kappa)=u_{n}(x,\kappa) \exp(i \kappa x)$, where the periodic part
$u_{n}(x,\kappa)$ of the Bloch state can be expanded as a Fourier
series
\begin{equation}
u_n(x, \kappa)= \frac{1}{\sqrt{2 \pi}} \sum_{l=-\infty}^{\infty}
\theta_{n,l}(\kappa) \exp[(2 i \pi l /a)x]
\end{equation}
with coefficients $\theta_{n,l}(\kappa)$. Owing to the orthogonal
and normalization conditions of Bloch states, the matrix
$\mathcal{A} \equiv \theta_{n,l}$ turns out to be unitary, i.e.
$\mathcal{A}^{-1}=\mathcal{A}^{\dag}$. In place of the Fourier
coefficients $\theta_{n,l}(\kappa)$, defined in the first
Brillouin zone $-\pi/a < \kappa  \leq \pi/a$, one can introduce
the set of functions
\begin{equation}
B_n(k=\kappa+2 \pi l/a)=\theta_{n,l}(\kappa)
\end{equation}
which are defined over $-\infty < k < \infty$. A typical example
of refractive index profile and corresponding band diagram
 is shown in Figs.1(b) and 1(c) for an excitation wavelength $\lambda=633$ nm,
 lattice period $a=11 \; \mu$m, bulk refractive index $n_s=1.43$ and
maximum index change of the waveguides $\Delta n=0.0016$. The
behavior of the Fourier coefficients $|B_n(k)|^2$ of Bloch states
is also depicted in Fig.1(d).\\
In presence of a transverse index gradient $F$, small enough to
neglect interband transitions (i.e. ZT), the solution to Eq.(1)
for an assigned beam distribution $\phi(x,0)$ at the $z=0$ input
plane can be written as a superposition of wave packets
$\phi_n(x,z)$ belonging to the different bands of the arrays which
propagate independently each other and undergo BOs, namely one can
write \cite{Longhi06}
\begin{equation}
\phi(x,z)=\sum_{n=1,2,3,..} \phi_n(x,z).
\end{equation}
For negligible ZT, the evolution of the wave packet $\phi_n$ can
be calculated in a closed form by using the acceleration theorem
\cite{Callaway74} and reads \cite{Longhi06}
\begin{equation}
\phi_n(x,z)= \int_{-\pi /a}^{\pi /a} d \kappa c_n(\kappa)
\varphi_n(x, \kappa+Fz/ \lambdabar) \exp[-i \gamma_n(\kappa,z)].
\end{equation}
In Eq.(6), the coefficients $c_n(\kappa)$ are determined by the
field distribution illuminating the array at the input plane
according to
\begin{equation}
c_n(\kappa)=\langle \varphi_n(x,\kappa)| \phi(x,0)\rangle=
\int_{-\infty}^{\infty} dx \varphi^{*}_n(x,\kappa) \phi(x,0)
\end{equation}
whereas the phase term $\gamma_n(\kappa,z)$ is given by
\begin{equation}
\gamma_n(\kappa,z)= \frac{1}{F} \int_{0}^{Fz / \lambdabar}d
\kappa' E'_n(\kappa+\kappa'),
\end{equation}
where
\begin{equation}
E'_n(\kappa)=E_n(\kappa)- F \frac{2 \pi}{a} \int_0^a dx \; i
u_n(x,\kappa) \frac{\partial u_n (x,\kappa)}{\partial \kappa}.
\end{equation}
is the band dispersion curve, corrected to account for a possible
topological (Berry phase) contribution \cite{Zak,note1}. In
particular, after a full BO cycle, i.e. for a propagation distance
$z=z_B$, one has $\phi_n(x,z_B)=\phi_n(x,0) \exp(-i \gamma_n)$,
where
\begin{equation}
\gamma_n=\frac{1}{F} \int_{0}^{2 \pi /a} d \kappa E'_n(\kappa).
\end{equation}
Since in the general case the phases $\gamma_n$ accumulated by the
various wave packets $\phi_n$ undergoing BOs  are not the same or
do not differ by multiplies of $2 \pi$, the output field
distribution $\phi(x,z_B)$ does not generally reproduce the input
one $\phi(x,0)$. The output field is in turn given by the
interference of the various wave packets $\phi_n(x,0)$ with the
appropriate phase delays $\gamma_n$ given by Eq.(10). This kind of
interference is discussed in details in the next section and may
lead to a Mach-Zehnder-like interferometry which does not require
ZT.

\section{Beam break up and recombination: Classical interference}
Let us consider a broad input beam, with narrow angular spectrum
and near-field distribution $G(x)$, impinging the array at the
incidence angle $\theta$, so that $\phi(x,0)=G(x) \exp(i \kappa
x)$ where $\kappa= n_s \theta / \lambdabar=(\pi/a) (\theta /
\theta_B)$ and
\begin{equation}
\theta_B=\frac{\lambda}{2an_s}
\end{equation}
is the Bragg angle \cite{note0}. We denote such an input field
distribution as $g^{(l)}(x,0;\kappa_0)$ and the corresponding
propagated field as $g^{(l)}(x,z; \kappa_0)$, where $\kappa_0$ and
the integer $l$ are defined such  that $\kappa_0+2 \pi l/a=\kappa$
and $-\pi/a \leq \kappa_0 <\pi/a$. The normalization condition
$\int dx |g^{(l)}(x,z; \kappa_0)|^2=1$ is assumed. In this case,
following the analysis of Ref.\cite{Longhi06}, one can show that
inside the arrayed structure one has
\begin{equation}
|\phi_n(x,z)|^2= \left| \theta_{n,l}(\kappa_0) \varphi_n \left(
x,\kappa_0+\frac{Fz}{\lambdabar} \right)G(x-x_n(z)) \right|^2
\end{equation}
 where we have set
\begin{equation}
x_n(z)=\frac{1}{F} \left[ E'_n(\kappa_0+Fz/
\lambdabar)-E'(\kappa_0) \right].
\end{equation}
Equations (5) and (12) indicate that the injected beam $\phi(x,0)$
breaks into a superposition of wave packets $\phi_n$ belonging to
the different bands of the array, with weighting factors
$|\theta_{n,l}(\kappa_0)|^2$, which undergo BOs along different
paths $x_n(z)$ [see, for instance, Figs.3(a) and 4(a)-(d) to be
discussed below]. According to Eq.(13), the path $x_n(z)$ followed
by the wave packet $\phi_n$ reproduces the shape of the band
dispersion curve $E_n(\kappa)$, eventually corrected to include
the Berry phase contribution.
 After a full BO cycle, i.e. at
$z=z_B$, the different wave packets $\phi_n$ interference and the
following scattering relations hold (see the Appendix)
\begin{equation}
g^{(l)}(x,z_B; \kappa_0)=\sum_{\rho=-\infty}^{\infty}
\mathcal{M}_{l,\rho} g^{(\rho)}(x,0; \kappa_0)
\end{equation}
where the scattering matrix $\mathcal{M}$ is given by
\begin{equation}
\mathcal{M}=\mathcal{A}^{\dag} \mathcal{B} \mathcal{A}, \;
\mathcal{A}_{n,l}=\theta_{n,l}(\kappa_0), \; \mathcal{B}_{\rho,n}=
\exp(-i \gamma_n) \delta_{\rho,n},
\end{equation}
and $\gamma_n$ are given by Eq.(10). Note that, if the differences
of phase delays $\gamma_n-\gamma_{\rho}$ were multiplies of $2
\pi$, a full reconstruction of the input beam -apart from an
unimportant phase term- would be achieved since
$\mathcal{A}^{\dag}\mathcal{A}=\mathcal{I}$. In this case the beam
leaving the array at $z=z_B$ would not split and would diffract at
the same tilting angle $\theta$ as that of the incoming beam.
However, the phase delays $\gamma_n$ do not generally satisfy the
previous condition, and thus the beam leaving the array at $z=z_B$
 breaks up into several beams which diffract at the different
angles $\theta+2\rho \theta_B$ ($\rho=0, \pm 1, \pm 2, ...$)
according to Eq.(14). After some propagation distance, such beams
are not overlapped and can be thus spatially resolved. This
general behavior is schematically illustrated in Fig.2(a). More
generally, we can say that, if the array is simultaneously excited
by a set of identical broad beams tilted at different angles
$\theta+2 l \theta_B$ ($l=0, \pm 1, \pm2,...$), the array behaves
like a liner multiport optical system for the amplitudes of modes
$g^{(l)}$ [see Fig.2(b)] with a scattering matrix $\mathcal{M}$
given by Eqs.(15). Here we have focused our analysis to the case
where the length $L$ of the array equals the BO cycle $z_B$.
Similar results are obtained by assuming -more generally- that the
array length $L$ comprises an integer number $N$ of BO cycles,
i.e. for $L=Nz_B$. In this case, the scattering matrix can be
again factorized as $\mathcal{M}=\mathcal{A}^{\dag} \mathcal{B}
\mathcal{A}$, where now $\mathcal{B}_{\rho,n}=\exp(-i
N \gamma_n) \delta _{\rho,n}$.\\
As an important example, let us consider the case where only the
 two lowest bands of the array are involved in the dynamics, and the scattering relations (15)
may be limited to two modes solely. An inspection of the curves
$E_n(\kappa)$ and $|B_n(k)|^2$ shown in Figs.1 (c) and (d)
indicates that this
 condition is realized, to a good approximation, when the array is excited
  by two broad beams tilted the former at an angle $\theta_1$ equal to
 or slightly smaller than the Bragg angle $\theta_B$, the latter
at the angle $\theta_2=\theta_1-2 \theta_B \sim -\theta_B$ [see
Fig.3(a)]. If we use the simplified notations $g_1(x,z) \equiv
g^{(0)}(x,z;\kappa_0)$ and $g_2(x,z) \equiv
g^{(-1)}(x,z;\kappa_0)$ for the two modes scattered by the array
in one BO cycle, where $\kappa_0 \sim \pi/a$, the input-output
relations [Eq.(14)] reduce to
\begin{eqnarray}
g_1(x,z_B) & = & S_{11}g_1(x,0)+S_{12}g_2(x,0) \\
g_2(x,z_B)& = &S_{21}g_1(x,0)+S_{22}g_2(x,0)
\end{eqnarray}
\begin{figure}[htbp]
\includegraphics[width=8.4cm]{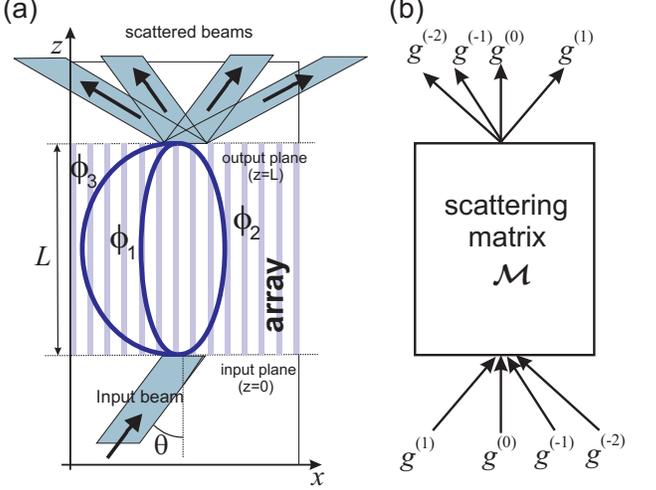}\\
   \caption{(color online) (a) Schematic of BO motion in the multiband regime. A broad beam illuminates the array at the incidence angle $\theta$ and
   excites several wave packets
   $\phi_1$, $\phi_2$, $\phi_3$ belonging to the different bands of the array. The wave packets undergo BOs along
   distinct  paths which reproduce in real space the spectral shapes of the band dispersion curves. After one BO cycle,
   owing to different phase delays $\gamma_n$ accumulated by the wave packets, several beams, diffracted at
   the angles $\theta$, $\theta \pm 2 \theta_B$, $\theta \pm 4 \theta_B$,
   ...., are produced. (b) Schematic of a linear-optic multiport system described by a
   scattering matrix $\mathcal{M}$.} \label{fig2}
\end{figure}
where we have set $S_{11}=\mathcal{M}_{0,0}$,
$S_{12}=\mathcal{M}_{0,-1}$, $S_{21}=\mathcal{M}_{-1,0}$ and
$S_{22}=\mathcal{M}_{-1,-1}$. According to Eqs.(15), the $2 \times
2$ scattering matrix $S$ entering in Eqs.(16) and (17) is given by
\begin{equation}
S= \left(
\begin{array}{cc}
\rho_{11}^* & \rho_{21}^* \\
\rho_{12}^* & \rho_{22}^*
\end{array} \right)
\times \left(
\begin{array}{cc}
\exp(-i\gamma_1) & 0 \\
0 & \exp(-i\gamma_2)
\end{array} \right)
\times \left(
\begin{array}{cc}
\rho_{11} & \rho_{12} \\
\rho_{21} & \rho_{22}
\end{array} \right)
\end{equation}
where
\begin{equation}
\left(
\begin{array}{cc}
\rho_{11} & \rho_{12} \\
\rho_{21} & \rho_{22}
\end{array} \right)= \left(
\begin{array}{cc}
\theta_{1,0}(\kappa_0) &  \theta_{1,-1} (\kappa_0)  \\
  \theta_{2,0}(\kappa_0) & \theta_{2,-1}(\kappa_0)
\end{array} \right).
\end{equation}
Since the matrix $\rho_{ik}$ is unitary, one has
$|\rho_{11}|=|\rho_{22}|$, $|\rho_{12}|=|\rho_{21}|$,
$|\rho_{11}|^2+|\rho_{12}|^2=1$ and $\rho_{11}
\rho_{12}^*+\rho_{21} \rho_{22}^*=0$. Without loss of generality,
we may assume $\rho_{11}=\rho_{22}$ to be real-valued and positive
by a suitable choice of the absolute phases of $u_{1}(x,\kappa_0)$
and $u_2(x,\kappa_0)$. After setting $\rho_{11}=\rho_{22}=
\sqrt{T}$ and $R=1-T$, where
$T=|\theta_{1,0}(\kappa_0)|^2=|\theta_{2,-1}(\kappa_0)|^2$, we can
then write
\begin{equation}
\left(
\begin{array}{cc}
\rho_{11} & \rho_{12} \\
\rho_{21} & \rho_{22}
\end{array} \right)=\left( \begin{array}{cc}
\sqrt{T} & \sqrt {R} \exp(i \alpha) \\
-\sqrt{R} \exp(-i \alpha) & \sqrt{T}
\end{array} \right)
\end{equation}
where $\alpha$ is the phase of $\rho_{12}$. Note that Eq.(20)
 is analogous to the scattering matrix of a lossless beam splitter with
transmittance $T=1-R$ (see, for instance, \cite{Campos89}).
Physically, such a transformation corresponds to the mixing of the
incoming beams $g_1$ and $g_2$ into the two wave packets $\phi_1$
and $\phi_2$ belonging to the two lowest-order bands of the array
 [see Fig.3(a)]. At exact Bragg
incidence, i.e. for $\theta_1=-\theta_2=\theta_B$, one has
$T=1/2$, i.e. the equivalent beam splitter is balanced. Note also
that the full transformation (18) is analogous to that of a
two-port Mach-Zehnder interferometer in which two waves $g_1$ and
$g_2$ are mixed by a fist beam splitter BS1, undergo different
phase delays $\gamma_1$ and $\gamma_2$ in the two arms of the
interferometer, and are then recombined by a second beam splitter
BS2 [see Fig.3(b)]. The transfer matrix S of the equivalent
Mach-Zehnder interferometer reads explicitly
\begin{widetext}
\begin{equation}
\left(
\begin{array}{cc}
S_{11} & S_{12} \\
S_{21} & S_{22}
\end{array} \right)=
\left(
\begin{array}{cc}
T \exp(-i \gamma_1) +R \exp(-i\gamma_2) &  \sqrt{RT} \exp(i \alpha) \left[\exp(-i \gamma_1)-\exp(-i \gamma_2) \right]\\
 \sqrt{RT} \exp(-i \alpha) \left[\exp(-i \gamma_1)-\exp(-i \gamma_2)
 \right] & T \exp(-i \gamma_2) +R \exp(-i\gamma_1)
\end{array} \right).
\end{equation}
\end{widetext}
Classical Mach-Zehnder interferometry, which do not require ZT, is
thus expected to be observable in multiband BOs. As an example,
for single beam excitation at the tilt angle $\theta_1$, i.e. for
$\phi(x,0)=g_1(x,0)$, the fractional light powers $P_1(\Delta
\gamma)$ and $P_2(\Delta \gamma)$ of the two beams $g_1(x,z_B)$
and $g_2(x,z_B)$ leaving the array and measured by two
photodetectors D1 and D2 [see Fig.3(a)], are readily calculated
from Eqs.(16), (17) and (21) as
\begin{eqnarray}
P_1(\Delta \gamma) & = & |S_{11}|^2=T^2+R^2+2RT \cos (\Delta
\gamma) \\
 P_2(\Delta \gamma) & = & |S_{21}|^2=2RT \left( 1-\cos
(\Delta \gamma) \right).
\end{eqnarray}
According to Eq.(10), the phase difference $\Delta \gamma$ is
given by
\begin{equation}
\Delta \gamma = \frac{2 \pi}{F a}  \left( \frac{a}{2 \pi}
\int_0^{2 \pi /a} d \kappa \left[ E^{'}_2(\kappa)-E^{'}_1(\kappa)
\right] \right) \equiv \frac{2 \pi \Delta E}{F a}
\end{equation}
\begin{figure}[htbp]
\includegraphics[width=8.4cm]{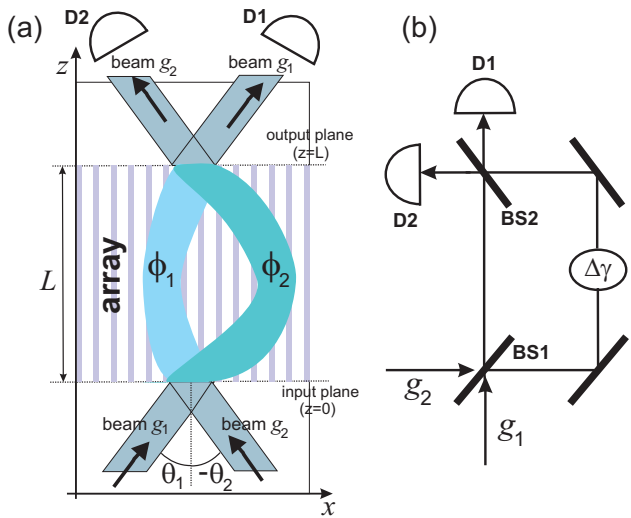}\\
   \caption{(color online) (a) Schematic of two-band BOs in a waveguide array excited by two input broad beams tilted
   at the angles $\theta_1 \sim \theta_B$ and $\theta_2=\theta_1-2 \theta_B \sim -\theta_B$.
   The paths followed by the wave packets $\phi_1$ and $\phi_2$ undergoing BOs map in real space the spectral shapes
   of the band dispersion curves $E_1(\kappa)$ and $E_2(\kappa)$, respectively. (b) Equivalent two-port Mach-Zehnder interferometer.}\label{fig3}
\end{figure}
where $\Delta E$ is the distance between the first and second
bands of the array, measured from their dc values \cite{note1}.
Typically the value $\Delta E$ is of the order of the refractive
index change $\Delta n$ of the core region of waveguides from the
dielectric substrate [see Fig.1(c)]. In practice, to vary $\Delta
\gamma$ one can slightly change the index gradient $F$ around the
reference value $F_0=\lambda/(aL)$ \cite{note2}. In this way, the
deviation of $z_B=\lambda/(Fa)$ from $L=\lambda/(F_0a)$ is
negligible, i.e. the BO cycle is almost completed at the output of
the arrayed region, whereas the change of $\Delta \gamma$ can be
 of the order of $2 \pi$ or larger as $Fa$ is typically much larger than $\Delta E$.
 Figures 4(a-d) show typical examples of two-band BOs
observed in numerical simulations of Eq.(1) for the
one-dimensional array of Fig.1. The length of the arrayed region
is $L=23$ mm, and the condition $z_B=L$ is attained for an applied
refractive index gradient $F_0=2.502 \; {\rm m}^{-1}$. For such a
relatively low value of refractive index gradient,  ZT from band 1
to band 2, and from band 2 to band 3, turns out to be negligible.
The array is excited by a broad Gaussian beam tilted at the Bragg
angle $\theta_2=-\theta_B$, and the evolution of beam intensity
$|\phi(x,z)|^2$ along the sample is plotted for a few values of
the refractive index gradient $F$ close to $F_0$. Beam splitting
and beam recombination after a full BO cycle are clearly visible,
as well as the change of the power levels in the two scattered
output beams as the applied index gradient $F$ is slightly varied
(by a few percents) at around $F_0$. The behavior of the
fractional powers $P_1$ and $P_2$ carried by the two output beams
versus the ratio $F_0/F$ is depicted in Fig.4(e), clearly showing
an oscillatory behavior with a visibility of about $\sim 90 \%$.
\begin{figure}[htbp]
\includegraphics[width=8.6cm]{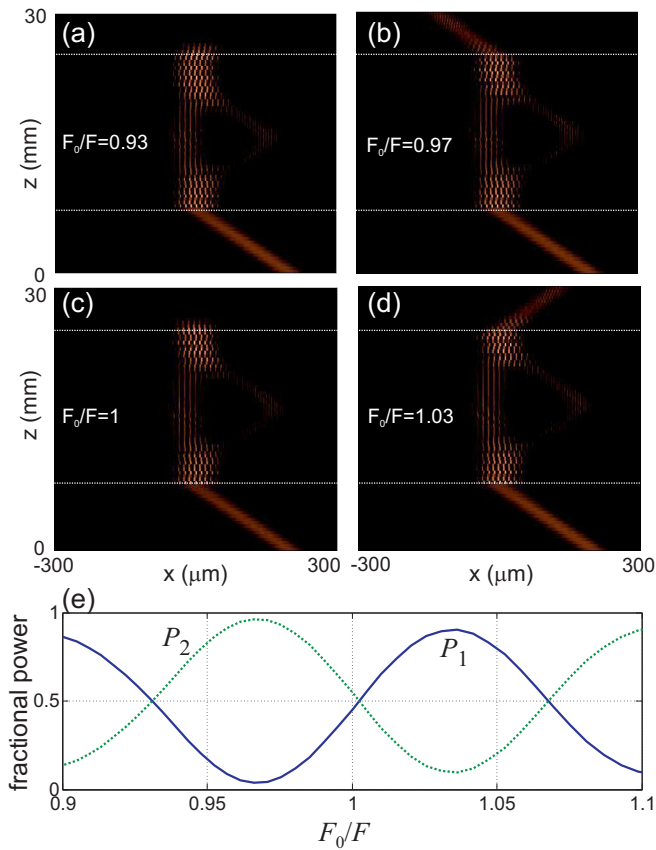}\\
   \caption{(color online) Classical Mach-Zehnder interferometry based on two-band BOs. (a),(b),(c) and (d) show
   the evolution of light intensity distribution along the optical structure, excited at the input plane by a broad Gaussian beam tilted
   at $\theta_2=-\theta_B$, for a few values of $F_0/F$. The arrayed structure has a length $L=23$ mm
   and it is comprised between the two dotted horizontal lines.
   The index gradient $F_0$, at which the array length $L$ is
   exactly equal to one BO cycle $z_B$, is given by $F_0= 2.502 \; {\rm
   m}^{-1}$. In (e) the numerically-computed fractional powers $P_1$ and $P_2$ carried
   by the two output beams (right- and left-diffracted beams, respectively) are plotted as functions of the ratio
   $F_0/F$.}\label{fig4}
\end{figure}
\section{Quantum interference in multiband Bloch oscillations with nonclassical light}
In previous sections, multiband optical BOs and interference
effects have been investigated in the framework of classical
electromagnetic theory. Here we extend the previous analysis to
nonclassical states of light undergoing multiband BOs, and discuss
relevant quantum interference effects similar to those observed in
Mach-Zehnder-based quantum interferometry
\cite{Rarity90,Boto00,Edamatsu02,Walther04}. A second-quantization
approach to study optical BOs in waveguide arrays, based on a
procedure similar to that adopted in the quantum theory of
solitons \cite{Haus89}, has been developed in Ref.\cite{Longhi08}.
In this approach, the scalar wave equation for the classical field
$\phi(x,z)$ [Eq.(1)] is written in Hamiltonian form assuming the
paraxial spatial coordinate $z$ as an independent variable, and a
quantization procedure is then applied by introducing creation
$\phi^{\dag}(x)$ and annihilation $\phi(x)$ bosonic field
operators (for details see \cite{Longhi08,Longhi09}). In the
Schr\"{o}dinger picture, the quantum field is described by a
quantum state $|\mathcal{Q}(z)\rangle$ which can be expanded in
Fock space as
\begin{equation}
|\mathcal{Q}(z) \rangle=\sum_{n=1}^{\infty} a_n |f^{(n)}\rangle,
\end{equation}
 where the $n$-photon number state $|f^{(n)} \rangle$ is
defined by
\begin{eqnarray}
|f^{(n)} \rangle  & = & \int d x_1 dx_2 ... dx_n
\frac{f^{(n)}(x_1,x_2,... x_n,z)}{\sqrt{n!}}
  \times \nonumber \\
  & \times & \hat{\phi^\dag}(x_1)  \hat{\phi^\dag}(x_2) ...
\hat{\phi^\dag}(x_n) |0\rangle.
\end{eqnarray}
The normalization conditions $\sum_n|a_n|^2=1$ and $\int dx_1 dx_2
... dx_n |f^{(n)}(x_1,x_2,...,x_n,z)|^2=1$ are assumed. The Fock
state $|f^{(n)} \rangle$ defined by Eq.(26) is obtained from the
vacuum state $|0\rangle$ by creating $n$ photons with a spatial
weighting function $f^{(n)}(x_1,x_2,...,x_n,z)$. The evolution
equation for $f^{(n)}$, obtained from the Schr\"{o}dinger equation
and using the commutation relations of field operators, reads
\cite{Longhi08}
\begin{equation}
i \lambdabar \frac{\partial f^{(n)}}{\partial z}=\sum_{l=1}^{n}
\left[ -\frac{\lambdabar^2} {2 n_s}
 \frac {
\partial^2}{\partial x_l^2}+V(x_l)-Fx_l \right] f^{(n)}.
\end{equation}
Owing to the bosonic nature of photons, solely symmetric functions
$f^{(n)}$ should be considered. The simplest $n$-photon number
state, denoted by $|g\rangle_n$, is obtained by assuming
$f^{(n)}=g(x_1,z)g(x_2,z)... g(x_n,z)$, where the function
$g(x,z)$ satisfies the classical wave equation (1) with the
normalization $\int dx |g(x,z)|^2=1$. Note that in this case one
has
\begin{equation}
|g\rangle_n=\frac{1}{\sqrt {n!}} \left( \int dx g(x,z)
\hat{\phi^\dag}(x) \right)^n |0 \rangle,
\end{equation}
so that this quantum state describes the excitation of the array
with a beam with a spatial profile $g(x,0)$ at the entrance plane
$z=0$ and carrying $n$ photons. More generally, for a given set of
orthogonal and normalized functions $g_1(x,z)$, $g_2(x,z)$, ...
that satisfy Eq.(1), one can construct the $n$-photon number state
$| g_1,g_2,...\rangle_{n_1,n_2,...}$ defined by
\begin{widetext}
\begin{equation}
| g_1,g_2,...\rangle_{n_1,n_2,...} = \frac{1}{\sqrt {n_1!}} \left(
\int dx g_1(x,z) \hat{\phi^\dag}(x) \right)^{n_1} \times
\frac{1}{\sqrt {n_2!}} \left( \int dx g_2(x,z) \hat{\phi^\dag}(x)
\right)^{n_2}  \times .... |0 \rangle,
\end{equation}
\end{widetext}
with $n_1+n_2+...=n$. Such a quantum state describes the
excitation of the array with a set of independent beams each
carrying $n_1$, $n_2$,... photons ($n=n_1+n_2+...$). The classical
picture of multiband BOs, described in previous sections, is
retrieved from the quantum model when the input beam is in a
coherent state (classical light), i.e. when a superposition of
Fock states with a Poisson distribution is considered
\cite{Longhi08}. Here we focus to the case where the input beam
describes a nonclassical field. In particular, let use suppose
that the array is excited by a set of tilted beams $g^{(l)}(x,0;
\kappa_0)$ ($l=0, \pm 1, \pm 2,...) $, which have been introduced
in Sec.III, and assume that $g^{(l)}$ is in a photon number state
(a Fock state) carrying $n_l$ photons. After introduction of the
creation operators $\hat{a}^{\dag}_l$
\begin{equation}
\hat{a}^{\dag}_l = \int dx g^{(l)}(x,0;\kappa_0)
\hat{\phi}^{\dag}(x),
\end{equation}
the quantum state of the system at the input plane of the array
may be written as \begin{eqnarray} |\mathcal{Q}(z=0)\rangle & = &
\frac{1}{\sqrt{... n_0! n_1! n_2 ! ....}}... \hat{a}_{0}^{n_0}
\hat{a}_{1}^{n_1}
\hat{a}_{2}^{n_2}...|0 \rangle \nonumber \\
& \equiv & |..., n_0,n_1,n_2,...\rangle.
\end{eqnarray}
Note that, as $g^{(l)}(x,0; \kappa_0)$ are orthogonal functions,
the operators $\hat{a}^{\dag}_l$ and $\hat{a}_l$ satisfy the
commutation relations
\begin{equation}
[\hat{a}_l,\hat{a}^{\dag}_{\rho}]=\delta_{l, \rho} \; \; , \;
[\hat{a}_l,\hat{a}_{\rho}]=[\hat{a}^{\dag}_l,\hat{a}^{\dag}_{\rho}]=0.
\end{equation}
 The quantum
state of the system at the output plane, i.e. after one BO cycle,
can be derived from Eqs.(14) and (29), and reads explicitly
\begin{widetext}
\begin{equation}
|\mathcal{Q}(z=z_B)\rangle  =  \frac{1}{\sqrt{... n_0! n_1! n_2 !
....}}... \left( \sum_{\rho} \mathcal{M}_{0,\rho} \hat{a}_{\rho}
\right)^{n_0} \left( \sum_{\rho} \mathcal{M}_{1,\rho}
\hat{a}_{\rho} \right)^{n_1} \left( \sum_{\rho}
\mathcal{M}_{2,\rho} \hat{a}_{\rho} \right)^{n_2} ...|0 \rangle .
\end{equation}
\end{widetext}
Note that, since the multiband BOs problem over one BO cycle
admits of a formulation in terms of a scattering matrix [Eq.(14)],
the quantum state at the output plane, as given by Eq.(33), is
consistent with the result that one would obtain using the
input-output operator formalism commonly adopted for linear
quantum-optical networks (see, for instance, \cite{Leonhard03}).
Equation (33) is at the basis of quantum interference and quantum
entanglement observable in multiband BOs when the array is excited
by photon number states. As an example, we discuss in detail the
doubling of interference fringes in photon counting rates for
correlated photon pairs undergoing two-band BOs, a phenomenon
analogous to the doubling of interference fringes in photon
correlation measurements observed in a Mach-Zehnder interferometer
and attributed to the so-called de Broglie wavelength of the
entangled biphoton state (see, for instance, \cite{Edamatsu02}).
To this aim, let us consider the two-band BO problem of Fig.3(a)
discussed in Sec.III. Instead of using classical light (i.e.
coherent states for the two beams $g_1$ and $g_2$), let us
illuminate the array by correlated photon pairs generated by
frequency-degenerate spontaneous parametric down-conversion and
incident onto the array at the Bragg angles $\theta_1=\theta_B$
and $\theta_2=-\theta_B$. The quantum state of light at the
entrance plane of the array can be thus written as
\begin{equation}
|\mathcal{Q}(z=0)\rangle=|1\rangle_{1}|1\rangle_{2},
\end{equation}
 where $|n_1\rangle_1 |n_2 \rangle_2$ denotes a $n=n_1+n_2$ photon number
state with $n_1$ photons in the mode $g_1$ and $n_2$ photons in
the mode $g_2$. Using Eqs.(16), (17) and (34), from Eqs.(31) and
(33) it readily follows that the state of quantum field after one
BO cycle is given by
\begin{widetext}
\begin{equation}
|\mathcal{Q}(z=z_B)\rangle= \sqrt 2
S_{11}S_{21}|2\rangle_{1}|0\rangle_{2}+ \left(
S_{11}S_{22}+S_{12}S_{21} \right) |1\rangle_{1}|1\rangle_{2}+\sqrt
2 S_{12}S_{22}|0\rangle_{1}|2\rangle_{2}.
\end{equation}
\end{widetext}
The joint probabilities $R_{11}$ and $R_{22}$  to find both
photons in the same beam, either $g_{1}$ or $g_2$, after one BO
cycle are then given by
\begin{equation}
R_{11}=2 |S_{11}|^2 |S_{21}|^2 \; , \; \;  R_{22}=2 |S_{12}|^2
|S_{22}|^2.
\end{equation}
Such relations give the two-photon counting rates that one would
measure in an experiment \cite{Edamatsu02}. Using Eq.(21) and
assuming $T=R=1/2$ (valid for incidence angles exactly tuned at
the Bragg angles $\pm \theta_B$), one finally obtains
\begin{equation}
R_{11}=R_{22}= \frac{1}{4} \left( 1- \cos (2 \Delta \gamma)
\right).
\end{equation}
 Note that the counting rates $R_{11}=R_{22}$ oscillate
like $\sim \cos(2 \Delta \gamma)$, i.e. at {\it twice} the phase
difference $\Delta \gamma$, as opposed to the classical
first-order interference fringes \cite{note3}, which oscillate
like $\sim \cos(\Delta \gamma)$ [see Eqs.(22) and (23)]. The
doubling of the counting rate oscillation frequency versus the
phase delay $\Delta \gamma$ is analogous to that observed in an
ordinary Mach-Zehnder interferometer probed by correlated photon
pairs and generally explained as a manifestation of the Broglie
wavelength of the biphoton entangled state formed after the first
beam splitter BS1 and probed at the second beam splitter BS2 of
the interferometer \cite{Edamatsu02}.

\section{Conclusions}
In this work interference phenomena for classical and
non-classical light propagating in arrays of coupled waveguides
and undergoing multiband optical Bloch oscillations with
negligible Zener tunneling have been theoretically investigated. A
wave scattering analysis of multiband BOs shows that
Mach-Zehnder-like interference effects spontaneously arise owing
to beam splitting and subsequent beam recombination occurring at
each BO cycle. A noteworthy example of quantum interference is
provided by the doubling of the interference fringes in photon
counting rates for a correlated photon pair undergoing two-band
BOs. This phenomenon is analogous to the one observed in a
Mach-Zehnder interferometer excited by pairs of correlated photons
and is a manifestation of the so-called Broglie wavelength of the
two-photon entangled state produced after the first beam splitter
of the interferometer and probed by the second one
\cite{Edamatsu02}. It is envisaged that the present results may
stimulate further theoretical and experimental investigations of
classical and quantum interference phenomena of light propagating
in complex periodic, quasi-periodic or disordered photonic
lattices.

\appendix
\section{Derivation of the scattering matrix} In this appendix
we derive the scattering relations given in the text [Eqs.(14) and
(15)] for the amplitudes $g^{(l)}$ of tilted beams at the input
and output planes of the array, i.e. after one BO cycle. To this
aim, let us first prove Eqs.(14) and (15) when the incident beams
are tilted plane waves, i.e. let us first assume
\begin{equation}
g^{(l)}(x,0;\kappa_0)= \frac{1}{\sqrt{2 \pi}} \exp \left( i
\kappa_0 x +i \frac{2 \pi}{a}l x \right)
\end{equation}
($l=0, \pm 1, \pm 2, ...$). Owing to the completeness of Bloch
states $\varphi_{n}(x,\kappa)$, one can write
\begin{equation}
g^{(l)}(x,0;\kappa_0)= \sum_{n=1,2,3,...} \int_{-\pi/a}^{\pi/a} d
\kappa \langle \varphi_n | g^{(l)} \rangle \varphi_{n}(x,\kappa).
\end{equation}
Since $\varphi_n(x,\kappa)=u_n(x,\kappa) \exp(i \kappa x)$ and
taking into account the Fourier decomposition of $u_n(x,\kappa)$
[Eq.(3)], one readily finds $\langle \varphi_n(x,\kappa)|
g^{(l)}(x,0; \kappa) \rangle =\theta_{n,l}^{*}(\kappa_0)
\delta(\kappa-\kappa_0)$, so that Eq.(A2) yields
\begin{equation}
g^{(l)}(x,0;\kappa_0)= \sum_{n=1,2,3,...} \mathcal{A}^{\dag}_{l,n}
\varphi_n(x,\kappa_0)
\end{equation}
where we have introduced the matrix $\mathcal{A}_{n,l} \equiv
\theta_{n,l}(\kappa_0)$ and $\mathcal{A}^{\dag}$ is the adjoint of
$\mathcal{A}$, i.e.
$\mathcal{A}^{\dag}_{l,n}=\mathcal{A}^{*}_{n,l}$. After one BO
cycle, the Bloch state $\varphi_n(x,\kappa_0)$ accumulates a phase
shift $\exp(-i \gamma_n)$, where $\gamma_n$ is given by Eq.(10).
After the introduction of the diagonal matrix
$\mathcal{B}_{\rho,n}=\exp(-i \gamma_n) \delta_{n,\rho}$, we can
thus write
\begin{eqnarray}
g^{(l)}(x,z_B;\kappa_0) & = &  \sum_{n=1,2,3,...}
\mathcal{A}^{\dag}_{l,n} \exp(-i \gamma_n)\varphi_n(x,\kappa_0)
\nonumber \\
& = & \sum_{n=1,2,3,...} (\mathcal{A}^{\dag}\mathcal{B})_{l,n}
\varphi_n(x,\kappa_0).
\end{eqnarray}
Since the matrix $\mathcal{A}$ is unitary,
$\mathcal{A}^{-1}=\mathcal{A}^{\dag}$ and Eq.(A3) can be inverted
yielding
\begin{equation}
\varphi_n(x,\kappa_0)= \sum_{l=0, \pm 1, \pm 2,...}
\mathcal{A}_{n,l} g^{(l)}(x,0;\kappa_0).
\end{equation}
Substitution of Eq.(A5) into Eq.(A4) finally yields
\begin{equation}
g^{(l)}(x,z_B;\kappa_0) = \sum_{\rho=0,\pm 1, \pm 2,...}
(\mathcal{A}^{\dag}\mathcal{B} \mathcal{A})_{l, \rho}
g^{(\rho)}(x,0;\kappa_0)
\end{equation}
which are precisely the scattering relations between input and
output waves given in the text [Eq.(14)], with a scattering matrix
$\mathcal{M}=\mathcal{M}(\kappa_0)$ given by
$\mathcal{M}=\mathcal{A}^{\dag} \mathcal{B} \mathcal{A}$. Such
relations can be extended to the case where the input waves
$g^{(l)}(x,0; \kappa_0)$ are not strictly plane waves, rather
tilted broad beams with a narrow angular spectrum $\hat{G}(\Delta
\kappa)= (2 \pi)^{-1/2} \int dx G(x) \exp(-i \kappa x)$ and
near-field distribution $G(x)$. In this case we may write
\begin{eqnarray}
g^{(l)}(x,0;\kappa_0) & = & \frac{1}{\sqrt {2 \pi}} \int d \Delta
\kappa \hat{G}(\Delta \kappa) \times \nonumber \\
& \times & \exp \left(i \kappa_0 x +i \Delta \kappa x+ i \frac{2
\pi}{a}l x \right).
\end{eqnarray}
Repeating the previous analysis to each of the plane waves
entering in the integral on the right hand side of Eq.(A7) and
using the superposition principle yields
\begin{widetext}
\begin{equation}
g^{(l)}(x,z_B;\kappa_0) =  \sum_{\rho=0, \pm 1, \pm 2,...} \int d
\Delta \kappa \frac{1}{\sqrt{2 \pi}} \hat{G}(\Delta \kappa)
\mathcal{M}_{l,\rho}(\kappa_0+\Delta \kappa) \exp \left( i
\kappa_0x+i\Delta \kappa x+i \frac{2 \pi}{a} \rho x \right).
\end{equation}
\end{widetext}
If the Fourier coefficients $\theta_{n,l}(\kappa)$ of Bloch states
- and hence the transfer matrix $\mathcal{M}(\kappa)$- vary slowly
over the spectral extension of $\hat{G}(\Delta \kappa)$, we may
take $\mathcal{M}_{l,\rho}(\kappa_0+\Delta \kappa) \simeq
\mathcal{M}_{l,\rho}(\kappa_0) $ out of the integral in Eq.(A8).
The remaining terms left under the integral then yields precisely
$g^{(\rho)}(x,0;\kappa_0)$ [see Eq.(A7)]. We thus finally obtain
\begin{equation}
g^{(l)}(x,z_B;\kappa_0) =  \sum_{\rho=0, \pm 1, \pm 2,...}
\mathcal{M}_{l,\rho}(\kappa_0) g^{(\rho)}(x,0;\kappa_0)
\end{equation}
which is extends the scattering matrix formalism to the case of
broad beam excitation.


\begin{thebibliography}{}
\bibitem{note1}
For a symmetric refractive index profile of the waveguide, such as
for the lattice of Fig.1(b), the periodic part $u_n(x,\kappa)$ of
Bloch states can be chosen such that $E^{'}_n(\kappa)=E_n(\kappa)$
and the Berry phase correction to $\gamma_n$ vanishes.
\bibitem{note0}
The tilt and Bragg angles $\theta$ and $\theta_B$  are here
measured for light waves propagating inside the dielectric medium
of refractive index $n_s$ rather than in vacuum, as in
Ref.\cite{Longhi06}.
\bibitem{note2}
In polymer waveguide arrays, tuning of the index gradient $F$ can
be realized by thermal gradient control, as demonstrated in the
experiments of Refs.\cite{Morandotti99,Trompeter06}. In
lithium-niobate waveguide arrays, such as those used in the
experiment of Ref.\cite{Longhi06}, electro-optic tuning of the
phase difference $\Delta \gamma$ can be also achieved.
\bibitem{note3}
The classical interference fringes, oscillationg like $\sim \cos
(\Delta \gamma)$, are retrieved when the quantum state of the
input field is in a coherent state, or when a single photon, in
either one of the two beams $g_1$ or $g_2$, undergoes a two-band
BO cycle, i.e. for either $|\mathcal{Q}(z=0)\rangle=|1\rangle_1
|0\rangle_2$ or $|\mathcal{Q}(z=0)\rangle=|0\rangle_1
|1\rangle_2$.
\end{thebibliography}
\end{document}